\documentclass[journal]{vgtc}                % final (journal style)
\usepackage{mathptmx}
\usepackage{graphicx}
\usepackage{times}

\usepackage{mathtools}

\usepackage{float}

\usepackage[colorlinks=true,linkcolor=blue]{hyperref}%

\onlineid{0}

\vgtccategory{Research}

\vgtcinsertpkg

\title{$E^3$: Visual Exploration of Spatiotemporal Energy Demand}

\author{Junqi Wu, Zhibin Niu*, Jing Wu, Xiufeng Liu, Jiawan Zhang}
\authorfooter{
 Junqi Wu, Zhibin Niu, Jiawan Zhang are with College of Intelligence and Computing, Tianjin University. Jing Wu is with Cardiff University. Xiufeng Liu is with Technical University of Denmark.
 Corresponding: zniu@tju.edu.cn
}

\shortauthortitle{Biv \MakeLowercase{\textit{et al.}}: Global Illumination for Fun and Profit}
\abstract{Understanding demand-side energy behaviour is critical for making efficiency responses for energy demand management. We worked closely with energy experts and identified the key elements of the energy demand problem including temporal and spatial demand and shifts in spatiotemporal demand. To our knowledge, no previous research has investigated the shifts in spatiotemporal demand. To fill this research gap, we propose a unified visual analytics approach to support exploratory demand analysis; we developed  $E^3$, a highly interactive tool that support users in making and verifying hypotheses through human-client-server interactions. A novel potential flow based approach was formalized to model shifts in energy demand and integrated into a server-side engine. Experts then evaluated and affirmed the usefulness of this approach through case studies of real-world electricity data. In the future, we will improve the modelling algorithm, enhance visualisation, and expand the process to support more forms of energy data.} % end of abstract


\CCScatlist{ % not used in journal version
 \CCScat{K.6.1}{Management of Computing and Information Systems}%
{Project and People Management}{Life Cycle};
 \CCScat{K.7.m}{The Computing Profession}{Miscellaneous}{Ethics}
}

  \teaser{
 \centering
 \includegraphics[width=0.9\linewidth]{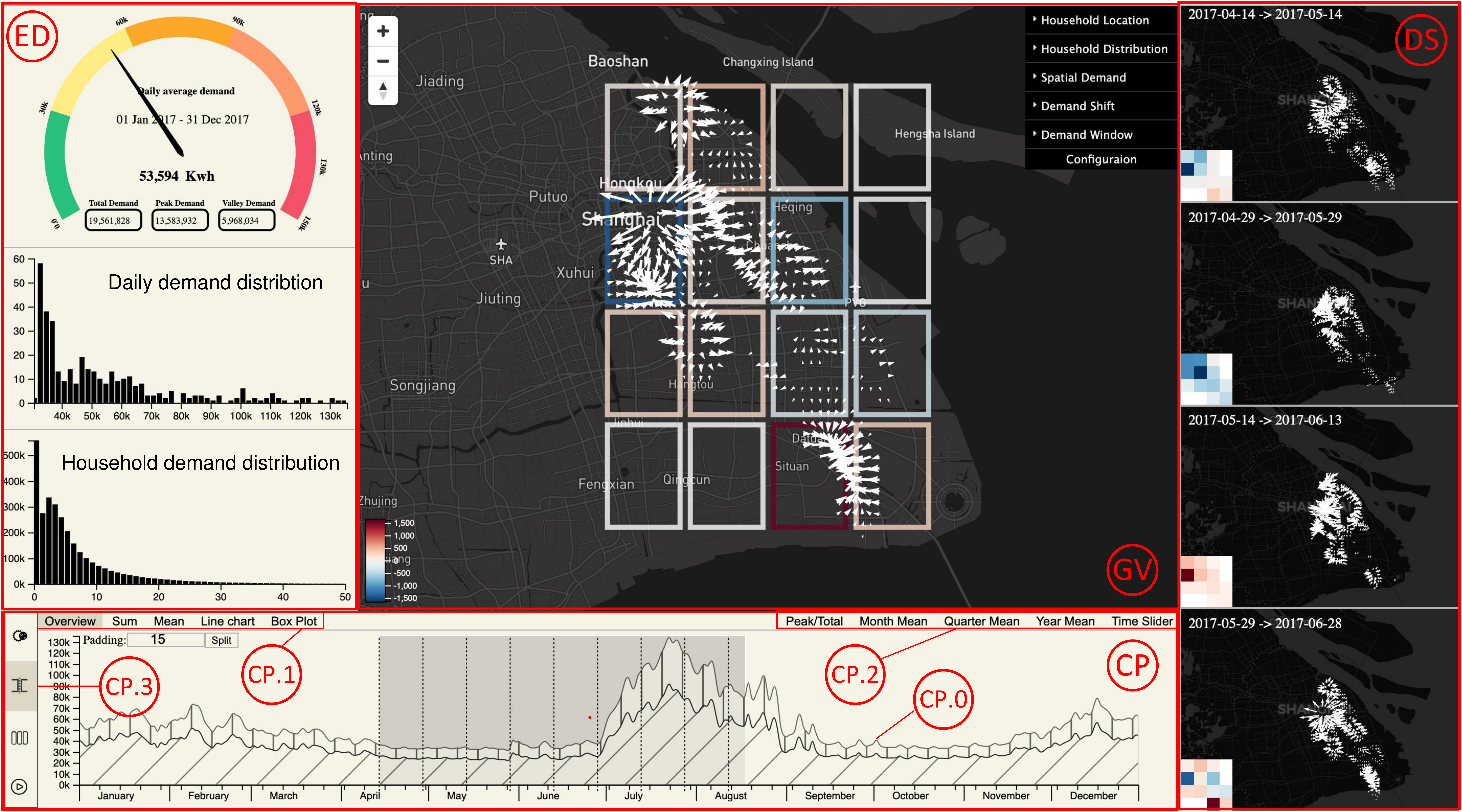}\vspace{-8pt}
  \caption{The system interface of $E^3$. Control Panel (CP) is composed of the temporal energy demand (CP.0) and a series of exploration tools (CP.1--CP.3). With these a user is able to interactively explore the temporal demand (CP.0), spatial demand ( see Geographical View (GV) ), and generate a spatio-temporal demand-shift task, viewing the results through the Demand-shift visual index (DS) for a quick selection and the Geographical view (GV) for fine details. Last, the Energy Demand Meter View (ED) quantifies the energy demand in Kilowatt hours and other values.
  }\vspace{-8pt}
  \label{interface}
  }

\begin{document}

%% The ``\maketitle'' command must be the first command after the
%% ``\begin{document}'' command. It prepares and prints the title block.

%% the only exception to this rule is the \firstsection command
\firstsection{Introduction}

\maketitle

\label{sec:intro}

In modern countries, Energy demand (or energy consumption) analysis is an important component in integrating energy planning and policy-making~\cite{bhatia1987energy, national1984improving}. Based on demand-side analysis, experts can make energy efficiency responses effectively (noted as energy demand management). For example, in the Peak-load Shifting or Peak Shaving operation, the load pressure of the power supply system can be effectively mitigated by advancing or delaying large energy load blocks through education or financial incentives~\cite{sun2013peak}. Although overall energy consumption may not necessarily be reduced, one could expect to decrease the demand for investment in networks and power plants. Energy demand behavioural research can improve the understanding of consumption patterns and be useful for demand management, even impacting relevant policy decisions~\cite{national1984improving}. However, in many cases, consumer preferences can not be reduced to an ideal socioeconomic metric and thus will be challenging to implement. The rapid growth of fine-grained smart meter data, data analytic technologies, and computing capability provide the opportunity to use data-driven energy demand monitoring, analyzing, and predicting.

We worked closely with experts in the energy sector and conducted interdisciplinary research to better understand energy demand. Since experts usually have few or only vague hypotheses about energy demand, particularly its dynamics, we set a goal to develop an ancillary visual exploration tool to allow them to perform exploratory data analysis through visual interactions and to gain insights into energy demand management. Here, we propose  $E^3$ -- a novel visual analytics tool, to assist experts to \emph{explore} temporal energy demand, to \emph{explore} spatial energy demand, and to \emph{explore} the spatiotemporal energy demand-shift.

We summarize the main contributions of this research as follows:

\vspace{-10pt}
\begin{enumerate}\setlength{\itemsep}{-0.1cm}
\item We propose a visual analytic framework that enables experts to construct energy demand hypotheses, verify these hypotheses and gain insights through an interactive human-client-server analysis loop.
\item We propose a novel potential flow based approach to model the energy demand-shift and integrate it into the server-side engine.
\item We develop $E^3$, a visual analytics tool with novel visualisations and interactions, enabling exploratory analysis.
\end{enumerate}
\vspace{-10pt}

%The rest of the paper is organized as follows: Section 2 describes works involving different aspects related to our problem; Section 3 details the background and requirement analysis of our approaches; Section 4 describes the methods and system. In Section 5, We present the results of case studies and expert feedback to evaluate the system. Discussions and conclusions are described in Section 6, Section 7.

\section{Related Work}

In this section, we first introduce the main data-driven visualisation and visual analytics research in energy demand management problems, then discuss recent progress on spatiotemporal visual analytics.

\emph{Data-driven energy demand analysis} research focuses on energy demand forecasting since the seminal concept proposed in 1984~\cite{national1984improving}. Amasyali and El-Gohary reviewed the data-driven building energy consumption prediction studies~\cite{AMASYALI20181192} and reported, among the machine learning algorithms used for energy forecasting, 47\% and 25\% of the studies utilized ANN and SVM, respectively, to train their models. In contrast, only 4\% of the studies utilized decision trees. 24\% of the studies utilized other statistical algorithms such as multiple linear regression, least squares regression, and auto regressive integrated moving average. Some recent progress in time series data such as LSTM, RNN are also widely applied to energy demand forecasting tasks. For example, CNN-LSTM neural networks are used to learn spatial and temporal features to predict housing energy consumption~\cite{KIM201972}.

Energy consumption pattern analysis also attracted research attention. For example, Hunt et al. model energy demand to analyse the trends and seasonal effects~\cite{HUNT200393}. Gaussian distributions and the Kullback-Leibler divergence-based clustering method are used to analyse household patterns of electricity consumption~\cite{6484217}. An association rule mining based quantitative analysis approach is designed for residential electricity consumption patterns analysis~\cite{wang2018association}. Occupant behavior has a major impact upon energy demand. Frequent pattern mining of human behavioral variations such as the high uncertainty about the order of use, varying time of use, and increased or reduced frequency of use of appliances, are employed in the energy consumption decision patterns mining~\cite{7894203}. Markov chains are extensively used to model occupant behavior and then to estimate energy demand and its fluctuations in the energy sector~\cite{KAVGIC20101683}. However,  Markov chains are limitations in accurately capturing the coordinated behavior of occupants and are prone to over-fitting. Rich features related to the coordination of occupants’ activities are employed to compare occupant to nearest-neighbor behavior~\cite{baptista2014accurate}.

\emph{Visual Analytics for Energy.} In the energy sector, prime visualisations such as graphs and bar charts are extensively used to provide comparable energy consumption data over time. The seminal interdisciplinary literature presents several power system visualisation techniques to help analyze the relationships between network power flows using techniques such as animation, contouring of bus and transmission line flow values, and interactive 3D data visualisation~\cite{overbye2000new}. Coincidence factor based heatmap visualisation is used to identify peak demand charges and avoid power outage~\cite{YARBROUGH201510}. Calendar-like pixel visualisations, with color boosting to anomaly scores, integrated with spiral visualisation, line chart, and tree maps that are designed to detect anomalies in power consumption data [16]. FigureEnergy is an interactive visualisation that allows users to annotate and manipulate a graphical representation of their electricity consumption data, and therefore make sense of their past energy usage by understanding when, how, and for what purpose, some amount of energy was used~\cite{costanza2012understanding}. Operational performance is integrated with building information modelling (BIM) as a visualisation dashboard to support the management of a building’s performance~\cite{gerrish2017bim}. Ambient and artistic visualisation for residential energy use feedback is explored, where Phyllotaxis design, Hive design, and Pinwheel design in energy use are discussed~\cite{rodgers2011exploring}. Matches, Mismatches, and Methods for Multiple-View Workflows for Energy Portfolio analysis are discussed~\cite{brehmer2015matches}. Mosaic Groups mapping encoded by household energy use combines with geodemographics to enable a better understanding energy user types in the UK~\cite{6400545}. GreenGrid is designed to explore the planning and monitoring of the Electricity Infrastructure. Geographic layout coming with a weighted network interface is designed to quickly identify where the system would be most likely to separate if an uncontrolled islanding event were to occur~\cite{4695829}.

\emph{Spatiotemporal Visualisation.} Spatiotemporal data visualisation has been extensively researched and applied in various domains~\cite{ANDRIENKO2003503,demvsar2015analysis}. A majority of the studies are trajectories analysis~\cite{Ferreira2013VisualEO,Lin2017ATA,Tominski2012StackingBasedVO}. Pattern extraction can be applied to obtain significant latent patterns from the movement data. Space-time cube representation is an information visualisation technique where spatiotemporal data points are mapped into a cube~\cite{4668344, Andrienko2013SpaceTF}. AirVis, is designed to assist domain experts to efficiently capture and interpret the uncertain propagation patterns of air pollution based on graph visualisations~\cite{Deng2019AirVisVA}. Multidimensional spatiotemporal data are modelled as tensors and then decomposed to extract the latent patterns for comparison and visual summarization~\cite{8440840}. Flow maps are employed to track the clustering behaviors and direction maps, drawn upon the orientation of vectors, can precisely identify the location of events~\cite{liu2019pattern}. Extending the density difference model, Kim et al. proposed a gravity-based flow extraction model, which can effectively separate human movement from spatiotemporal data without using trajectory information~\cite{kim2017data}. A population based vector field was proposed to visualise time-geographic demand momentum. By representing transport systems as vector fields that share the same time–space domain, demand can be projected onto the systems to visualize relationships between them~\cite{LIU201522}.

\section{DATA AND TASK ABSTRACTION}
This section introduces the background of our study, describes the relevant concepts and datasets, and summarizes the requirements.

\textbf{Problem Statement. }In the energy demand management practice, experts or decision makers always wish to quantify energy consumption, to understand location differences, and to understand consumer preferences by consumption behaviour analysis. However, most of the published work in this area focuses on analysing the energy consumption fluctuations over time in a certain area~\cite{suganthi2012energy}, and that does not meet domain requirements.

To address this issue, we conducted intensive discussions with two experts in the energy sector and categorised energy demand into Temporal Energy Demand (TED), Spatial Energy Demand (SED), and Spatiotemporal Energy Demand-shift (ST-EDS). TED refers to energy consumption fluctuation over a period of time in a certain area. The topic is extensively discussed in the energy sector, and the Peak-load Shifting mentioned above in the introduction section is part of the temporal energy demand management problem; SED is defined as the geospatial demand distribution across a certain area at a specific moment. ST-EDS is an important yet unexplored research problem that is introduced in this paper. It is defined as the spatiotemporal changes in energy consumption for a certain area during different periods. The ST-EDS can help discover energy consumption behaviour patterns, and thus lead to better demand management and planning.

\textbf{Summary of Analysis Tasks.} We summarize the following analysis tasks, incorporating the feedback from experts, to support our exploratory analysis. They are: \emph{[T.1] Holistically analyse the temporal and spatial energy demand.} A challenge for the experts to analyze the demand is the spatiotemporal nature of the energy consumption data. The entanglement of spatial and temporal distributions makes it difficult for energy experts to make and verify their hypotheses about the demand and its dynamic changes. In contrast to the vast traditional energy demand literature that discusses the two issues separately, an exploration that considers both temporal and spatial information comprehensively will have the potential to discover new knowledge. \emph{[T.2] Modelling the energy demand-shift through a suitable algorithm and support exploratory knowledge discovery.} The geospatial dynamics of energy consumption reflect the behaviour of household occupants and play a pivotal role for better demand management and planning. However, there are very few studies that address the issue. It requires modelling the demand-shift mathematically, and devising a user-friendly interface to support exploratory knowledge discovery. \emph{[T.3] Quantify the preceding exploration results.} Energy experts raised the requirement of quantifying the exploration results as numbers, is important for comparison and evaluation of the significance of the trends or patterns in the knowledge domain.

\textbf{Design Rationales.} Based on analysis tasks, we propose the following design rationales. \emph{D.1 Scalable, coordinated and appropriate temporal and spatial energy demand visualisation.} This requires creating a compact visualisation in a form familiar to domain experts’ mental simulations and follows Shneiderman’s seeking Mantra [28] (T.1, T.2). \emph{D.2 Highly interactive design to reinforce understanding. } An exploratory analysis requires intensive interactions to explore consumption data and to acquire knowledge about correlations between the various factors (T.1, T.2). \emph{D.3 Appropriate symbols and colour mapping for intuitive metaphors.} Intuitive representation is an essential element of most visualisation systems. Carefully designed visual metaphors can help experts reduce the visual burden and improve their understanding of the real situation. Visual mapping of the discovered patterns can facilitate visual sense-making and help to interpret the human behaviour behind the energy consumption behaviour (T.3).

\section{Method}
\label{sec:overview}
This section describes the method of the visual analytic system.

\subsection{Overview of the Approach}
\begin{figure*} [ht!]
  \centering
  \includegraphics[width=1\linewidth]{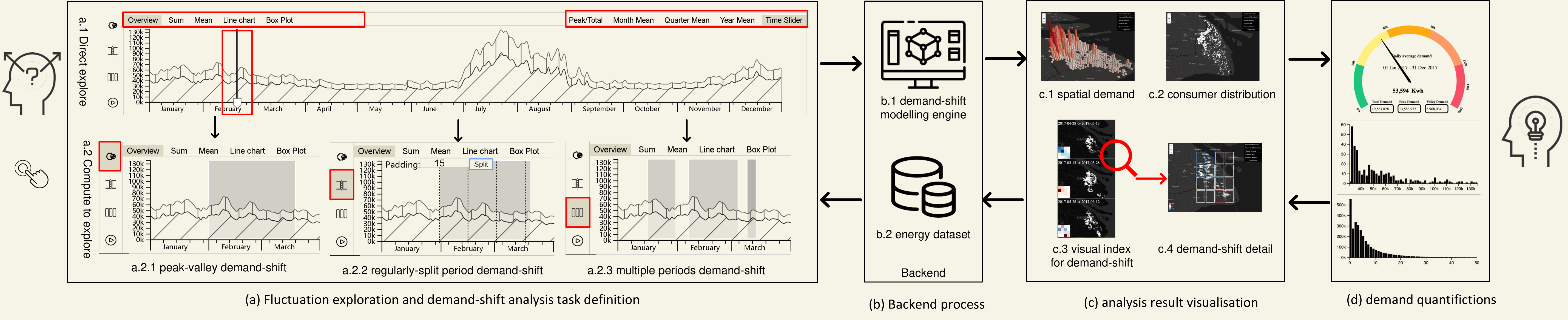}\vspace{-10pt}
  \caption{The visual analytics pipeline. Experts can start to explore the fluctuation demand (a.1) and then explore the spatial dimension (c.1 and c.2). The auxiliary lines (view CP.1 and CP.2) and demand quantifications (d) will help them to understand the spatial and temporal demand characteristics. With the initial exploration of temporal and spatial demand,  experts may already have some hypothesis about the demand-shift. They can formalise the hypothesis into analysis tasks (a.2.1, a.2.2, a.2.3) interactively and the interface will send the settings to the backend demand-shift engine (see Section 4.3). The computed results are sent back to the visual interface (View GV and DS in Fig.~\ref{interface}). The users can choose the demand-shift from the index view (c.3) and analyse the details in the detail view (c.4). In such a way, the users can construct the hypothesis, verify the idea, and thus gain insights through a human-client-server interactive analysis loop, interactively and iteratively.
  } \label{framework}\vspace{-18pt}
\end{figure*}

Motivated by the domain requirements, we designed and implemented $E^3$ to support an exploratory analysis. Fig.~\ref{framework} gives an overview of the approach. It consists of 1) front-end visual analytic interface, from which users may perform an exploratory analysis and define demand-shift tasks through interactions; 2) back-end server. It stores the data and performs demand-shift modeling based on the users' setting.

The control panel (CP) is usually the start point for demand fluctuation exploration and demand-shift analysis task definitions. As the step (a) in Fig.~\ref{framework} shows, users can directly explore the spatial demand for a certain area over a specific period (T.1), and define new demand-shift analysis tasks through interactions associated with the buttons on View CP.3 (see Fig.~\ref{interface}). Three kinds of analysis tasks are supported: the demand-shift of the peak-valley period, of a regular-split period, and of multiple periods. The tasks will be modelled and computed at the back-end engine (T.2) as specified in Section 4.3. The results will be visually summarized in the index view DS. Users are able to choose the result of interest and view the details on View GV.

\subsection{Data}
\begin{figure} [H]
  \centering
  \includegraphics[width=1\linewidth]{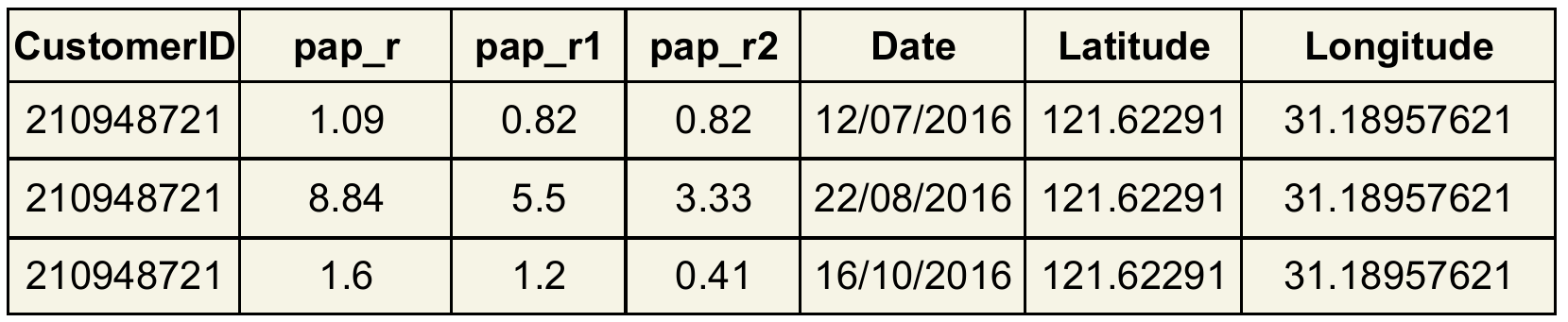}\vspace{-10pt}
  \caption{ The electrical household consumption data.}\label{data}\vspace{-12pt}
\end{figure}

The energy data are usually multiple-grained time serial records with geospatial information of the consumers. We use the electrical energy demand as a representation in this paper, more forms of energy data such as district heating will be supported in our extended work. We collected the electrical household consumption data from the Pudong area in Shanghai for a whole year. The electrical energy consumption data are usually daily-grained with peak and valley detailed time serial data. Fig.~\ref{data} shows sample rows of the data in seven columns. Among them, $pap\_r$ (Positive Active Power) gives the total electrical consumption of the day; $pap\_{r1}$ (Positive Active Power at Peak Time) gives the peak period (6:00 – 22:00), and $pap\_{r2}$ (Positive Active Power at Valley Time) gives the valley period (22:00 – 6:00) energy consumption value of the day.

\vspace{-5pt}

\subsection{Spatiotemporal Demand--shift Modelling}

As stated in Section 4.1, the users define the spatiotemporal demand-shift analysis tasks in the interface and send the settings to the backend engine for demand-shift modelling. In this section, we describe our potential flow-based modelling algorithm.

We observed that the geospatial energy demands change continuously over time, and thus the energy demand is a continuum occupying a simply-connected region in the time dimension with a irrotational characteristic. Inspired by the advancement of fluid dynamics and continuum mechanics, such a continuum can be modelled as a potential flow~\cite{batchelor2000introduction}. Formally, the spatiotemporal demand-shift is defined as $\nu = \bigtriangledown \varphi $. The spatiotemporal demand–shift is modelled as the flow velocity $\nu$, a vector field to describe the rate of demand change. It is defined as the gradient of the velocity potential $ \varphi $ that refers to the spatial energy demand change at the selected moments or periods.

In order to model the spatial energy demand, a kernel density estimation based approach is proposed to encode discrete household energy consumption into a continuous representation. Such an operation can efficiently generate a smoother vector field. In detail, it is modelled as $\hat{\phi}_{h}(x)=\sum_{i=1}^{n}c_{i}K_{H}{(x-x_{i})}$, where $x_i =$ the household locations, and $c_{i}  = $ the normalised energy demand to re-weight demand strength; $H = $ bandwidth (or smoothing), a $d \times d$ matrix, which is symmetrical and positive definite; $K = $ the kernel function. We chose to use the gaussian kernel in this work. The spatial energy demand change $\varphi$ can be modelled as the discrepancy between the estimated $\hat \phi$ of different moments.

\section{Visualisation and Interactions}

In this section, we describe the visual interface and interaction of the system. As Fig.~\ref{interface} shows, the system $E^3$ consists of four coordinated components. The panel view (CP) is usually the start point for temporal, spatial, and demand-shift analysis through highly interactions. As Fig.~\ref{framework} shows, the demand-shift analysis task will be modelled and computed at the backend and visualise the results in the view GV and DS ( Fig~\ref{interface}). In this section, we describe the key visual design details.

\subsection{Control Panel}

The control panel (View CP) displays the temporal energy demand (View CP.0) and provides the main functional interactions (View CP.2 and CP.3) to support spatial demand and demand-shift analysis.

In detail, the temporal energy demand view (CP.0) utilizes a stream graph to visualise the peak and valley energy demand. The user is able to toggle the auxiliary analysis line (yearly, quarterly, monthly average demand, and peak to valley ratio) in the View CP.2 to select the period of interest for further temporal analysis (T.1, T.2). The demand-shift (T.3) is started through the functional buttons in the view CP.3. After a user has a period of interest to analyse through the fore step, he or she can define the explore demand-shift tasks of different type of periods: peak-valley, regularly-split period, or multiple periods. The user can toggle the corresponding button and then select the period(s) of interest by brush operation on the View CP.0 and then toggle the compute button in View CP.3 to generate the results and list them in the Demand-shift visual index (view DS).

\subsection{Geographical View and Demand-shift Visual Index}

The geographical view GV enables users to view the spatial demand and demand-shift in the finest detail on a large map. It includes: 1) Spatial demand visualisation. A 3D hexagon visualisation (see Fig.~\ref{framework} c.1) is designed for the energy demand exploration (T.1), where the height is proportional to the energy demand and the colour depth is proportional to the household number. 2) Demand-shift visualisation. It gives the spatiotemporal demand-shift computation result from the backend engine. The Demand-shift Visual Index (View DS) is designed to give a visual summary and glance of the demand-shift generated by the exploration task (T.2). The view DS also works as a visual index for the users to quickly locate the demand-shift of interest and view the details on the Geographical View GV.

\begin{figure} [ht!]
  \centering
  \includegraphics[width=1\linewidth]{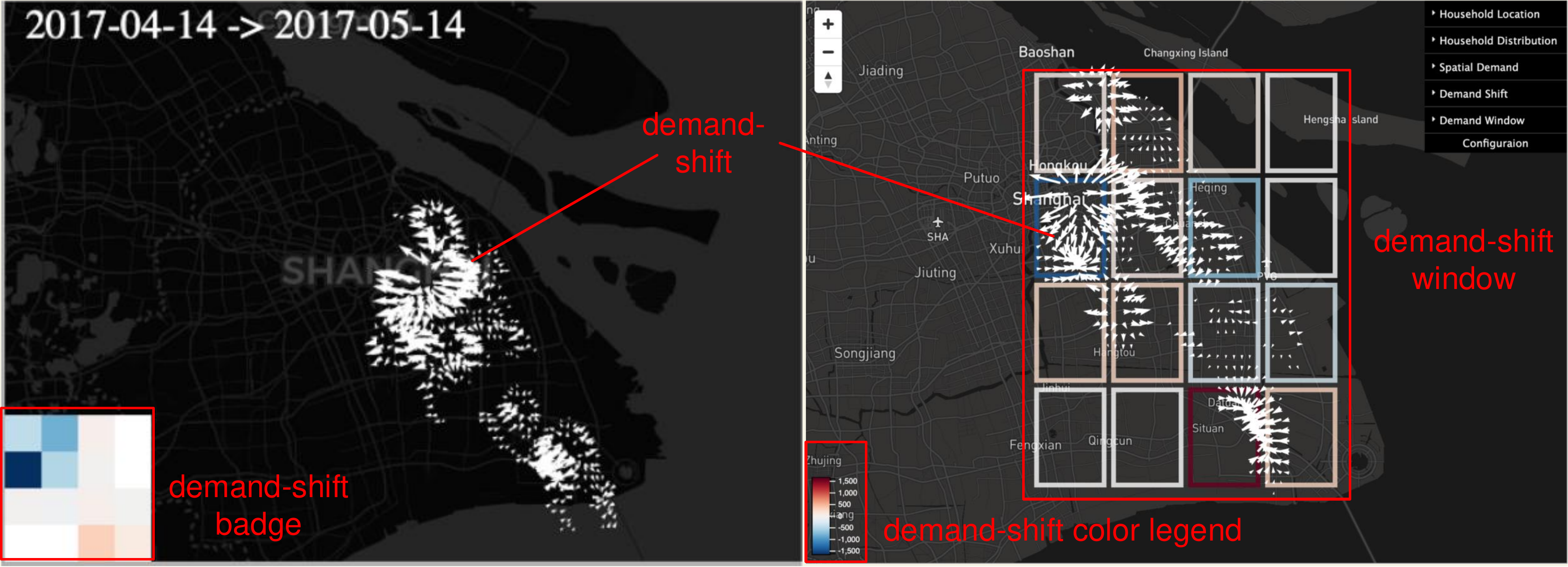}\vspace{-10pt}
  \caption{Visual design for energy demand-shift in the view GV and DS.}\label{design}\vspace{-12pt}
\end{figure}

We introduce four visual elements to represent the demand-shift, as shown in Fig.~\ref{design}: 1) a flow map is used to visualise the demand-shift where the length is encoded as the rate of demand change ($\nu$) ; 2) a demand-shift window on the geographical view gives the coverage of analysis and the colour is encoded as the spatial energy demand change $\varphi $. Such window-shape design does not obscure the map and gives the necessary quantification information for analysis; 3) a demand-shift color legend gives the corresponding absolute value of spatial energy demand change in the grid area. We use absolute value in response to the task T.3 for quantitative analysis. 4) a demand-shift badge also encodes the spatial energy demand change $\varphi $. It has the same meaning as the demand-shift window. It gives a visual summary of the demand change of the grided areas on the visual index view. We use a solid grid instead of the frame design as it is much smaller on the visual index; as well, a solid design is more prominent.

\vspace{-8pt}

\subsection{Energy Demand Meter View}

The energy demand meter view (View ED) gives the statistical overview of the energy demand and the distributions of the demand on a daily and household basis. The design of the meter is consistent with the charting used in the energy sector to give a quantified overview of the total, peak, valley, and average daily demand.

\section{Case Study}

The implemented system is evaluated by two energy experts. Expert A is a professor in Energy, who has over ten years of socioeconomic research in energy and Expert B is senior engineer working on the power system planning from the power supply company. We give a quick training on the core concepts and operations of the system. They work as a team to evaluate the system and we collected their feedback on the system usability, visual design, and limitations.

\vspace{-6pt}
\begin{figure} [H]
  \centering
  \includegraphics[width=1\linewidth]{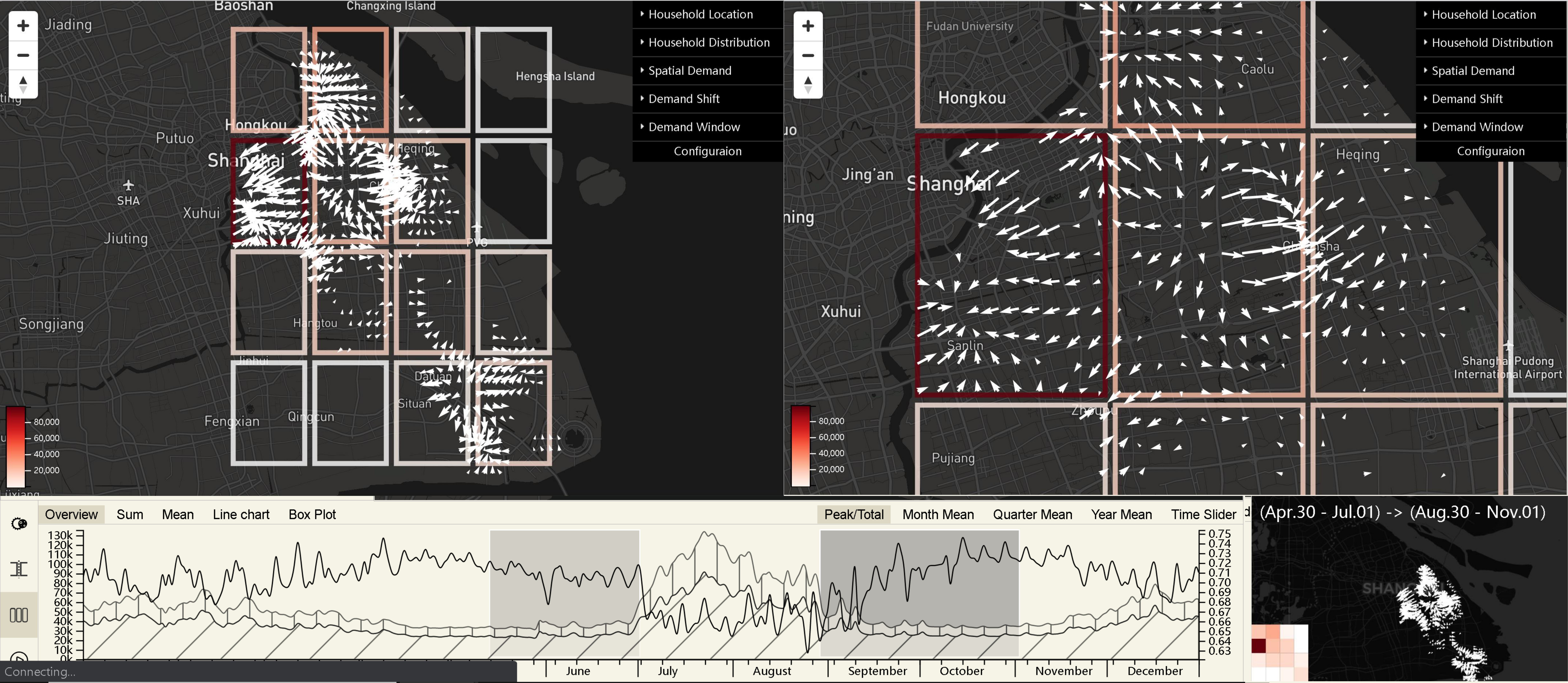}\vspace{-6pt}
  \caption{Investigating the demand-shift. Zoom in to view the legend.}\label{case}\vspace{-8pt}
\end{figure}

These experts started to explore the temporal energy demand fluctuation from the control panel, first observing an obvious bimodal pattern – a larger surge demands in the summer and a lower surge in the winter. This makes sense as people consume more electricity for cooling using air conditioners in the summer than heating in the winter in Shanghai. From the peak to total ratio curve, they observed: 1) A weekly periodic pattern which reflects the household behaviour pattern. 2) Much lower peak to total ratio in the summer than any other seasons. This means the power system is on full load and maintenance engineers should pay attention to any potential failures. Usually, the temperature in Shanghai is similar between May, June ($24^{\circ}$ -- $28^{\circ}$) and September, October ($27^{\circ}$ -- $23^{\circ}$), however, the experts found that the total energy demand in the later period is about 200,000 Kwh higher. The expert A explained that one of the causes is the psychological inertia and potential energy efficiency response people can be expected to take. The demand-shift attracted more of Expert B’s attention. From the demand-shift badge, he observed that the energy demands are unbalanced between regions.
The area in the top-right corner (Sanlin town, a highly dense residential area) has the largest spatial demand change (more than 80,000 Kwh). He zoomed in to view the flow map to observe the rate and direction of demand change. The flow map visualises an interesting crowd energy demand behavior pattern – the arrows give the convergence/divergence hotspot or lines. This means that such an area may change the load strongly; expert B suggested that a proper energy efficiency response, facility operation, maintenance and planning should be considered. Regarding feedback, expert B confirmed that the demand-shift as well as the temporal and spatial demand exploration are useful for understanding consumer behaviour and future power system planning. Both of them liked the compact visual designs. Expert A said that training is necessary as the semantic meaning of the demand-shift windows might be ignored. He also suggested that the current system can be more powerful if it supported more forms of energy data. For example, supporting to analysis district heating demand, a promising neat solution for the supply of low-carbon heat, has an important practical significance, he complemented.

\vspace{-10pt}
\section{Conclusions and Future Work}

We presented the initial development of an ongoing visualisation analytics study that explores the spatiotemporal energy demand of daily energy consumption. The main contributions at this stage include: characterising the problem of analyzing spatiotemporal energy demands and compiled visual exploration tasks, based on the iterative discussions with domain experts; a systematic visual analytic framework for temporal, spatial, spatiotemporal energy consumption behaviours is proposed; we developed a prototype system $E^3$ and had it evaluated by two experts. As we go forward, a more powerful potential flow algorithm will be created, which will generate better visualisation of spatiotemporal energy demands. The visual design will be modified to support multi-source and multi-granularity collected utilities data. Our collaborators plan to evaluate the system with more empirical data and use the prototype in several ongoing projects. %We hope the application of our interdisciplinary visual analytics research to the energy sector will help disseminate our results and achieve everyone’s energy efficiency target.

%% if specified like this the section will be committed in review mode
%\acknowledgments{
%This work was supported by NSFC (61802278), European Union Horizon 2020 Marie Skłodowska-Curie grant agreement (754462).
%
%}

\bibliographystyle{abbrv}
%%use following if all content of bibtex file should be shown
%\nocite{*}
%\bibliography{ref}

\end{document}